\documentclass[aps,pre,showpacs,floatfix,twocolumn,groupedaddress]{revtex4}

\usepackage{graphicx}
\usepackage{amsmath}
\usepackage{wasysym}
\usepackage{amssymb}
\usepackage{bm}

\begin{document}

\title{Dynamical Scaling Behavior of Percolation Clusters in 
Scale-free Networks}

\author{F. Jasch, Ch. von Ferber and A. Blumen}
\affiliation{Theoretische Polymerphysik, Universit\"{a}t Freiburg, 
Hermann-Herder-Str. 3, D-79104 Freiburg, Germany}

\date{\today}

\begin{abstract}
In this work we investigate the spectra of Laplacian matrices that
determine many dynamic properties of scale-free
networks below and at the percolation threshold.
We use a replica formalism to develop analytically, based on an
integral equation, a systematic way to determine the ensemble averaged
eigenvalue spectrum for a general type of tree-like networks. 
Close to the percolation threshold we find characteristic scaling
functions for the density of states $\rho(\lambda)$ of
scale-free networks. $\rho(\lambda)$  shows characteristic power laws 
$\rho(\lambda)\sim\lambda^{\alpha_1}$ or
$\rho(\lambda)\sim\lambda^{\alpha_2}$ for small $\lambda$, where 
$\alpha_1$ holds below and $\alpha_2 $ at the percolation threshold. 
In the range where the spectra
are accessible from a numerical diagonalization procedure the two
methods lead to very similar results. 
\end{abstract}

\maketitle

\section{Introduction}
Recent studies of nets, ranging from social networks to power grids
and the internet,
revealed that in many cases the degree distribution $p_k$, i.e. the
probability that an arbitrary vertex is connected to exactly $k$ other
vertices, often exhibits a power law, namely that $p_k\sim
k^{-\gamma}$ holds \cite{albert1,albert2}.    
Networks for which this relation is fulfilled are called {\em scale-free}; 
scale-free networks differ from the  
classical random graphs \cite{E73}, for which the distribution $p_k$ is 
Poissonian, and from small-world-networks 
\cite{WS98,NW99,SJB00,JB00,JB02,BJ02}. 
Recent works have clarified that the properties of scale-free networks,
in particular percolation, differ markedly from the classical case 
\cite{albert1,albert2,hav1,hav2,Hav}.
It turns out that the asymptotic behavior of $p_k$ for $k$ large, the
so-called tail of $p_k$, which is quantified by $\gamma$, is
fundamental in differentiating between the distinct classes of
behavior: Thus    
for $\gamma<4$  the critical exponents  change from
the usual values found for classical graphs \cite{Hav,CNSW}. 

Now, the topological properties of a network are
reflected in the spectral
properties of its connectivity matrix ${\bf C}$: 
This matrix is constructed by letting its off-diagonal elements 
$C_{ik}$ be $1$ if $i$ and $j$ are connected or $0$ otherwise; 
moreover, the diagonal elements $C_{ii}$ of ${\bf C}$ are zero.
For scale-free networks it was found that the
density of the eigenvalues of ${\bf C}$ has a triangular form with
a power-law tail \cite{GKK,FDBV,DGMS}. On the other hand, many problems 
ranging from the dynamics
of randomly branched polymers \cite{JFB03} and the stress relaxation of near
critical gels \cite{BAHZ01}, over random resistor-capacitor networks 
\cite{stephen} to glassy relaxation 
dynamics \cite{Br88}, depend on the Laplacian ${\bf A}$; ${\bf A}$ is
connected to ${\bf C}$ via: 
\begin{equation}
A_{ik}=\left(\delta _{ik}\sum\limits_{j=1}^{N}C_{jk}\right)-C_{ik}.  
\label{2}
\end{equation}
A whole series of works based on ${\bf A}$ were devoted to classical
deterministic and random graphs, such as Cayley-trees (dendrimers),
hyperbranched macromolecules, the
 Erd\"os-R\'enyi (ER) random graph and bond diluted Cayley-trees 
\cite{BarHaim97,Kopelman97,BAHZ01,BLMZ99,BLMZ00,stephen,Br88,JFB03,S74,BKB00,BKB01,FB,BJKF03}.
Based on the Laplacian, many time and frequency-dependent observables
can be written as integrals over $\rho(\lambda)$,
the density of eigenvalues  of ${\bf A}$: The structure of these
observables is either
\begin{equation}
  \label{dynint}
Q(t)=\int_{0+}^\infty d\lambda f(t,\lambda)\rho(\lambda), 
\end{equation}
or some related form, in which $t$ is replaced by $\omega$. For
instance, for random walks, the site
averaged return probability $Q_R(t)$ of a random walker to the origin 
is obtained with the
choice $f_R(t,\lambda)=e^{-\lambda t}$ \cite{Br88}.
Moreover, the mechanical storage and loss moduli \cite{BKB00,BKB01}, 
the averaged 
time-dependent stretching of macromolecules in external fields
\cite{BKB00,BKB01}, and the
dielectric relaxation functions \cite{GB02}, all obey forms similar to 
Eq. (\ref{dynint}). We are interested in the dynamic behavior
of random graphs with arbitrary degree distributions and hence in the
density $\rho(\lambda)$ of their eigenfrequencies.
Following the ideas used in the analysis of gel dynamics \cite{BAHZ01}
and hyperbranched polymers \cite{JFB03}, we display an integral
equation for $\rho(\lambda)$ for a special class of
random graphs with arbitrary
degree distributions \cite{M95,M98,CNSW,NSW}.
This integral equation allows us to determine  
$\rho(\lambda)$ for the 
classes of scale-free networks discussed in Ref. \cite{Hav}.

\section{Random graphs with arbitrary degree distributions}
The ensemble of networks under consideration is obtained by starting 
from $N$ vertices. Each vertex $i$ has its degree $k_i$, and the
probability distribution of the $k_i$ is $p_k$. As discussed in Ref. 
\cite{NSW}, one can then connect the vertices pairwise through bonds
(random pairing), while fulfilling 
the condition that the number of bonds emanating from each vertex $i$ 
is given by its degree $k_i$. All such possible combinations create
the ensemble.
In the limit $N\to\infty$  
the probability that a certain vertex is
involved in a closed loop of bonds vanishes like $1/N$ \cite{NSW};
thus, in this limit a typical network realization 
is a set of
connected  treelike clusters. Such a treelike structure may also be
created as follows: 
We start from a vertex, say $i$, whose random degree $k$ 
is chosen from the given distribution $p_k$. 
Then each of the $k_i$ bonds of  
vertex $i$ ends in a new 
vertex.
One must note now that the probability of reaching via a randomly
chosen bond a vertex of degree $k$ is proportional to $kp_k$,
i.e. it obeys the distribution 
\begin{equation}
q_k= \frac{k p_k}{\sum_{j=1}^\infty jp_j}.
\label{verbindprob}
\end{equation}  
It follows that we must now distribute the degrees of the newly
produced vertices according to $q_k$, Eq. (\ref{verbindprob}). The
procedure is then continued step after step and stops only where no
new bonds were produced in the previous step. The method creates all the
random trees of the ensemble. Two examples of such 
ensembles are the
bond diluted Cayley tree with functionality $f$ \cite{F61}, where 
$p_k={f \choose k} p^k(1-p)^{f-k}$, and the 
Erd\"os-R\'enyi random graph, whose degree distribution 
is $p_k=p^ke^{-p}/k!$ \cite{NSW}.

Depending on the $p_k$-distribution, the ensemble  
consists either exclusively of finite clusters or it includes
an infinite, connected cluster, containing a finite
fraction of the vertices of the system \cite{S85}.
In Refs. \cite{NSW,hav1} 
it was shown that the condition for the existence of this
infinite cluster (also called percolating cluster) is given by
\begin{equation}
  \frac{\sum_{k=0}^\infty k(k-1)p_k}{\sum_{k=0}^\infty kp_k}>1. 
\label{perkbed}
\end{equation}
Eq. (\ref{perkbed}) defines the so-called percolation threshold.
It is very useful to extend the present model, by also allowing
the strength of each bond to be weighted \cite{BAHZ01,JFB03} 
following a normalized
coupling strength distribution $D(\mu)$.
Thus in the corresponding connectivity matrices, each of the nonzero values of
$C_{ik}$ can be chosen according to the distribution $D(\mu)$.  

As mentioned above, for a given network cluster $S$ 
various dynamical quantities involve  
only the density of eigenvalues $\rho_S(\lambda)$ of 
the corresponding Laplacian ${\bf A}^S$. 
Now, the ensemble averaged density of eigenvalues is given by
\begin{equation}
\rho (\lambda )=\left\langle \rho _{S}(\lambda )\right\rangle \equiv
\sum_{S}w_{S}\rho _{S}(\lambda ),  
\label{averagedef}
\end{equation}
where  
the sum extends over all the clusters
$S$, each of the $\rho_S(\lambda)$ is normalized, and $w_S$ denotes the
probability with which the cluster $S$ is
produced by the iterative growth procedure.
Each of the $S$ created in this way is connected, so that ${\bf A}^S$ has only
one zero eigenvalue, whose corresponding eigenvector is homogeneous.
It turns out to be convenient to
split off from $\rho(\lambda)$ the delta peak at $\lambda=0$ whose 
weight is $\rho _{0}$, by setting: 
\begin{equation}
\rho (\lambda )=\rho _{0}\delta (\lambda )+\rho _{+}(\lambda ).  \label{6}
\end{equation}
The density of states is connected with the diagonal elements of the
resolvent, see e.g. Ref. \cite{JFB03}. Denoting for each site $k$ the
$k$-th diagonal element of the
resolvent ${\bf R}(\lambda)=({\bf A}^S-\lambda{\bf1})^{-1}$ by 
$R_{kk}(\lambda)$, what is needed is $R_{kk}(\lambda)$ 
averaged over all sites $k$ and over all the $S$-clusters:
\begin{equation}
  \label{R}
  R(\lambda)=\langle({\bf A}^S-\lambda{\bf1})^{-1}_{kk}\rangle.
\end{equation}
The probability $w_{k,S}$ 
that when creating $S$ we start at site $k$
does not depend on $k$;   
one has thus $w_{k,S}=w_S/|S|$, 
where $|S|$ denotes the number of vertices inside $S$. 
This leads to 
\begin{eqnarray}
R(\lambda ) &=&
\sum_S\sum_{k=1}^{|S|}w_{k,S}
({\bf A}^S-\lambda {\bf 1})^{-1}_{kk}
\nonumber\\
&=&
\sum_Sw_{S}\frac{1}{|S|}
\sum_{k=1}^{|S|}
({\bf A}^S-\lambda {\bf 1})^{-1}_{kk}.
\label{decomp}
\end{eqnarray}
Using for the normalized density of states of cluster $S$ the relation
\begin{equation}
\rho_S(\lambda)=\lim_{\epsilon \rightarrow 0}
\frac{1}{\pi}\frac{1}{|S|}\mbox{Im}\sum_{k=1}^{|S|}
({\bf A}^S-(\lambda+i\epsilon){\bf 1})^{-1}_{kk},
\label{rhos}
\end{equation}
we obtain from Eq. (\ref{averagedef})
\begin{equation}
\rho (\lambda )=
\lim_{\epsilon \rightarrow 0}\frac{1}{\pi }\mbox{Im}R(\lambda
+i\epsilon ), 
\label{densres}
\end{equation}
with $R(\lambda)$ being given by Eq. (\ref{R}).
Now, the average over the disorder can be performed with the help of
the replica method \cite{MPV86}.   

\section{Derivation of the integral equation}
In the following we denote the starting vertex by $0$.
To obtain the averaged trace of the resolvent we rewrite it 
with the help of a Gaussian integral over
$n$-dimensional vectors ${\bf r}_i$
\begin{widetext} 
\begin{equation}
R(\lambda )=\left\langle \left[ \text{Det}\frac{i({\bf A}^S-\lambda
      {\bf 1})}{%
2\pi }\right] ^{n/2}\frac{i}{n}\int \left( \prod\limits_{j}d{\bf r}%
_{j}\right) {\bf r}_{0}^{2}\exp \left[ -\frac{i}{2}\left(
\sum\limits_{j,k}A_{jk}^S{\bf r}_{j}{\bf r}_{k}-\lambda \sum\limits_{j}{\bf r}%
_{j}^{2}\right) \right] \right\rangle,  
\label{15}
\end{equation}
%\end{widetext}
see e.g. Ref. \cite{JFB03} for details.
The averaging procedure in Eq. (\ref{15}) is considerably simplified by
taking the replica limit $n\rightarrow 0$, since then the $n/2$-th power of the
determinant is unity.
Using
\begin{equation}
  \label{connection}
  \sum\limits_{i<j}C^S_{ij}({\bf r}_{i}-{\bf r}_{j})^{2}=
\sum\limits_{i,j}A^S_{ij}{\bf r}_{i}{\bf r}_{j},
\end{equation}
which follows readily from Eq. (\ref{2}), leads to 
%\begin{widetext}
\begin{equation}
R(\lambda) \dot{=} \frac{i}{n}\int 
\left( \prod\limits_{j}d{\bf r}_{j}\right) {\bf r}_{0}^{2}\exp \left[i\frac{
\lambda }{2}\sum\limits_{j}{\bf r}_{j}^{2}\right] \left\langle \exp
\left[-\frac{i}{2}
\sum_{j<k}C^S_{jk}\left( {\bf r}_{j}-{\bf r}_{k}\right) ^{2}\right] 
\right\rangle 
\label{R2}
\end{equation}
\end{widetext}
Here we use the dot over the equation sign to indicate that the limit 
$n\rightarrow 0$ has to be taken. 
Now we employ the fact that the $S$-clusters are trees, in order   
to perform the
integrations in Eq. (\ref{R2}) iteratively, following the number $g$ 
of growth steps.  
After the first growth step we have 
\begin{equation}
 R^{(1)}(\lambda)\dot{=}
\frac{i}{n}\int d{\bf r}_0\;
  {\bf r}_0^2\exp\left[i\frac{\lambda}{2}{\bf r}_0^2\right]
\sum_{k=0}^\infty p_k \{\phi^{(1)}({\bf r}_0)\}^k,  
\end{equation}
where we defined
\begin{equation}
 \phi^{(1)}({\bf r}_0)\equiv\int d{\bf r}\exp\left[i\frac{\lambda}{2}
{\bf r}^2\right]F({\bf r}_0,{\bf r})
\label{defphi1}
\end{equation}
and
\begin{equation}
 F({\bf r}_j,{\bf r}_k)=
\int_0^\infty d\mu \; D(\mu )\exp\left[-i\frac{\mu }{2}({\bf r}_j-{\bf r}_k)^2\right].
\label{defF}
\end{equation}
In a similar way, the averaged diagonal element after the second
growth step reads 
\begin{equation}
 R^{(2)}(\lambda)\dot{=}
\frac{i}{n}\int d{\bf r}_0\;
  {\bf r}_0^2\exp\left[i\frac{\lambda}{2}{\bf r}_0^2\right]
\sum_{k=0}^\infty p_k \{\phi^{(2)}({\bf r}_0)\}^k
\end{equation}
with
\begin{equation} 
\phi^{(2)}({\bf r}_0)=\int d{\bf r}\exp\left[i\frac{\lambda}{2}
{\bf r}^2\right]F({\bf r}_0,{\bf r})
\sum_{k=1}^\infty q_k\{\phi^{(1)}({\bf r})\}^{k-1}.  
\label{defphi2}
\end{equation}
More generally, introducing the generating functions 
of the $p_k$'s and the $q_k$'s: 
\begin{equation}
  G_0(\phi)=\sum_{k=0}^\infty p_k\phi^k\;\; \mbox{and}\;\;
  G_1(\phi)=\sum_{k=1}^\infty q_k \phi^{k-1}=\frac{
  G_0'(\phi)}{G_0'(1)},
\label{Gdef}
\end{equation}
we find that after $g$ growth steps $\phi^{(g)}({\bf r})$ obeys: 
\begin{equation}
 \phi^{(g)}({\bf r}_0)=\int d{\bf r}\exp\left[i\frac{\lambda}{2}
{\bf r}^2\right]
F({\bf r}_0,{\bf r})G_1[\phi^{(g-1)}({\bf r})],  
\label{rekphi}
\end{equation}
and that it can be obtained iteratively, starting from $\phi^{(1)}({\bf r})$
given by Eq. (\ref{defphi1}).
Furthermore, $R^{(g)}(\lambda)$ fulfills 
\begin{equation}
 R^{(g)}(\lambda) \dot{=}
\frac{i}{n}\int d{\bf r}_0\;
  {\bf r}_0^2\exp\left[i\frac{\lambda}{2}{\bf r}_0^2\right]
G_0[\phi^{(g)}({\bf r}_0)].  
\end{equation}
Now, the $n\to0$ limit can be performed as described in Ref.
\cite{JFB03}. This leads for $g\to\infty$ to the pair of equations 
\begin{equation}
\label{Rend}
R(\lambda)=-\frac{1}{\lambda}
\int_{0}^{\infty }dx\,e^{-x}G_0[\phi(x)] 
\end{equation}
and
\begin{equation} 
\phi(x)=\hat{{\bf O}}e^{-x}G_1[\phi(x)],  
\label{39}
\end{equation}
where ${\bf \hat{O}}$ is the linear operator 
\begin{eqnarray}
 \hat{{\bf O}}&=&\int_0^\infty d\mu
 D(\mu)\exp\left[-\frac{\lambda}{\mu}x\partial_x^2\right]
\nonumber\\
&=&\sum\limits_{k=0}^{\infty }\frac{\left\langle \mu
^{-k}\right\rangle_\mu}{k!}\left( -\lambda \right) ^{k}(x\partial
_{x}^{2})^{k},  \label{21}
\end{eqnarray}
where $\left\langle ...\right\rangle_\mu$ denotes the average over the
distribution $D(\mu)$.

In Ref. \cite{NSW} it was shown that the generating function 
$H_0(z)=\sum_{s=1}^\infty P_sz^s$ 
of the probabilities $P_s$ that a randomly chosen 
vertex is part of a cluster of $s$ vertices
can be obtained based on the relations 
\begin{equation}
H_0(z)=zG_0(H_1(z))\;\;\mbox{and} \;\; H_1(z)=zG_1(H_1(z)). 
\label{h0}
\end{equation}
Here $H_1(z)$ is the generating function for the distribution of sizes
of components that are reached by choosing a random bond and following
it to one of its ends.
As a check for our scheme we now show that our 
Eqs. (\ref{Rend}) and (\ref{39}) are consistent with Eqs. (\ref{h0}).
To do this we look for a solution of Eq. 
(\ref{39}) in the form of a power series in $\lambda$
\begin{equation}
  \label{ansatz}
  \phi(x)=\sum_{k=0}^\infty\lambda^k\phi_k(x).
\end{equation}
We obtain the first term by comparing powers of $\lambda$:
\begin{equation}
   \phi_0(x)=e^{-x}G_1(\phi_0(x)).
\label{phi0}
\end{equation}
Identifying $e^{-x}$ with $z$, Eq. (\ref{phi0}) reproduces 
the second Eq. (\ref{h0}) with 
$\phi_0(x)=H_1(e^{-x})$.
Hence we infer from the first Eq. (\ref{h0}) that 
\begin{equation}
  \label{interpret}
  e^{-x}G_0(\phi_0(x))=\sum_{s=1}^\infty P_s e^{-sx}.
\end{equation}
It follows that $G_0[\phi_0(0)]$ is the probability for a vertex to be part of 
a finite size cluster.  

From Eqs. (\ref{6}) and (\ref{densres}) one infers that 
$R(\lambda)$ possesses a simple pole of the form $\rho_0/\lambda$, where 
$\rho_0$ is the finite weight of zero eigenvalues. Now $\rho_0$  
can be calculated by inserting 
Eq. (\ref{ansatz}) into Eq. (\ref{Rend}), which leads to 
\begin{eqnarray}
\rho_0 &=&\int_0^\infty dx e^{-x}G_0(\phi_0(x))\nonumber\\
&=& 1+G_0'(1)\int_0^\infty dx 
e^{-x}G_1(\phi_0(x))\phi'_0(x)\\
&=&1+G_0'(1)\int_0^\infty dx 
\phi_0(x)\phi_0'(x)=1-\frac{G_0'(1)}{2},
\label{rho0}
\end{eqnarray}
where in the second step we performed a partial integration and used
Eq. (\ref{Gdef}).
We note that inserting Eq. (\ref{interpret}) into
(\ref{rho0}) leads to $\rho_0=\sum_{s=1}^\infty P_s/s$;
the result represents the fact that each 
$s$-cluster contributes a term $1/s$ to the density of the eigenvalue 
zero.

\section{Scale-free networks close to their percolation threshold}
In this section we turn from our general considerations to focus on 
{\em scale-free} degree distributions;
these exhibit for $k$ large a power-law behavior, 
$p_k\sim k^{-\gamma}$. To describe the distance from the percolation
threshold, Eq. (\ref{perkbed}), we introduce the parameter $\Delta$ through
the relation 
\begin{equation}
  \label{deldef}
\Delta=1-\frac{\sum_{k=0}^\infty k(k-1)p_k}{\sum_{k=0}^\infty k
p_k}=1-G_1'(1). 
\end{equation}
Evidently, we assume by this that the first and the second moments of
the $p_k$ distribution exist. 
From Eq. (\ref{perkbed}) it follows that for $\Delta>0$ the ensemble
is made up of
finite connected clusters, while for $\Delta<0$ there exists an
infinite cluster.
The critical point (percolation threshold) is at $\Delta=0$. 
As a note of caution we remark that the choice of the sign of 
$\Delta$ is possibly misleading but
since in the following we investigate
exclusively network clusters below the percolation threshold
this choice considerably simplifies the formulas.
Exemplarily, for the
ER random graph the critical point is at $p_c=1$ and for
the bond diluted Cayley tree it is at $p_c=1/(f-1)$; using Eq. 
(\ref{deldef}) it turns out that
in both cases $\Delta=(p_c-p)/p_c$. 
Note that for $\gamma<3$ Eq. (\ref{deldef})
diverges; this agrees with the criterium of Eq. (\ref{perkbed}), since
for $\gamma<3$ one always has an infinite cluster \cite{hav1}.
 
We center now on the form of Eqs. (\ref{39}) and (\ref{Rend})
close to $\Delta=0$.
To be sufficiently general, we assume $p_k$ to have for 
large $k$ the form
 \begin{equation}
p_k\sim k^{-\gamma}\{c+O(k^{-1})\}.
\label{scalefree}  
\end{equation}
This implies for $\gamma>3$ 
that the power series  of $G_1(\phi)$, Eq. (\ref{Gdef}),
has as radius of convergence the unit circle $|\phi|=1$. 
To determine the singularity on the radius of convergence
we remark that the expansion coefficients $\tilde{p}_k$ of the quantity
$\tilde{G}_1(\phi)\equiv G_1(\phi)-c\Gamma(2-\gamma)(1-\phi)^{\gamma-2}$ 
obey $\tilde{p}_k\sim k^{-\gamma-1}$ for large $k$.
Thus $\tilde{G}_1(\phi)$ is $m$-times continuously 
differentiable for $|\phi|\leq1$, 
where $m$ is the largest integer smaller than $\gamma-2$. 
Using the Taylor expansion of $\tilde{G}_1(\phi)$ around $\phi=1$ 
up to order $m$ one gets \cite{Bialas,Hav}
\begin{eqnarray}
G_1(\phi) &\simeq& 1+(1-\Delta)(\phi-1)+\ldots\\
\nonumber
&&+\frac{1}{m!}\partial_\phi^mG_1(1)(\phi-1)^m+
c\Gamma(2-\gamma)(1-\phi)^{\gamma-2},  
\label{gprop}
\end{eqnarray}
where we used Eq. (\ref{deldef}).
Close to the percolation threshold $\Delta=0$ we expect the solution
of Eq. (\ref{39}) to scale in its variables $x$ and $\lambda$, and we
choose a solution of the form \cite{stephen}
\begin{equation}
\phi(x)\simeq 1-\Delta^\beta\tilde{\phi}(x/\Delta^{\delta},
\lambda/\Delta^{1+\delta}), 
\label{scalean}
\end{equation}
with exponents $\beta>0$ and $\delta>0$, to be determined below. 
Inserting Eq. (\ref{scalean}) into Eq. (\ref{39}) and expanding
in powers of $\Delta$ by using Eq. (\ref{gprop}) we obtain  
\begin{eqnarray}
\lefteqn{1-\Delta^\beta\tilde{\phi}(x,\lambda)=}&&
\\&&
\Big\{1-\langle\mu^{-1}\rangle\lambda\Delta x\partial_x^2+\ldots\Big\}
\Big\{1-\Delta^\delta x+\ldots\Big\}\times
\nonumber\\&&
\Big\{1-
(1-\Delta)\Delta^\beta\tilde{\phi}(x,\lambda)+
\theta(\gamma-4)\frac{G_1''(1)}{2}\Delta^{2\beta}
\tilde{\phi}^2(x,\lambda)
\nonumber\\&&
+\theta(4-\gamma)c\Gamma(2-\gamma)\Delta^{\beta(\gamma-2)}
\tilde{\phi}^{\gamma-2}(x,\lambda)+\ldots
\Big\},\nonumber
\end{eqnarray}
where $\theta(x)$ denotes the Heaviside function and the dots indicate
terms with higher powers of $\Delta$. 
Comparing powers of $\Delta$ leads to the equation
\begin{eqnarray}
 0&=& 
\langle\mu^{-1}\rangle\lambda\Delta^{1+\beta}x\partial_x^2\tilde{\phi}(x)
-\Delta^{\delta}x+\Delta^{\beta+1}\tilde{\phi}(x) 
\nonumber\\&&
+\theta(\gamma-4)\frac{G_1''(1)}{2}\Delta^{2\beta}
\tilde{\phi}^2(x)
\nonumber\\&&
+\theta(4-\gamma)c\Gamma(2-\gamma)\Delta^{\beta(\gamma-2)}
\tilde{\phi}^{\gamma-2}(x).
\label{scalendiff}
\end{eqnarray}
Now the unknown exponents $\beta$ and $\delta$ are determined by the 
requirement that all terms in this equation be of the same order in $\Delta$. 
For $\gamma>4$ this leads to  $\beta=1$ and $\delta=2$. 
From Eq. (\ref{scalendiff}) it follows then:
\begin{equation}
 0=\langle\mu^{-1}\rangle\lambda x\partial_x^2\tilde{\phi}(x)
-x+\tilde{\phi}(x)+\frac{G_1''(1)}{2}\tilde{\phi}^2(x),  
\label{scale1}
\end{equation}
This universal scaling 
equation for the order parameter field $\tilde{\phi}(x)$ 
was already pointed out in Ref. \cite{stephen} in connection with
the mean-field theory of random resistor networks.
Thus for $\gamma>4$ we obtain the classical mean-field scaling
equation of the order parameter field $\phi(x)$, which is also valid
for the ER graph and the Cayley-tree.
This is in accordance with the result in Ref. \cite{Hav}, that 
the critical properties of classical 
random graphs are {\em not} changed for $\gamma>4$.
On the other hand, for $3<\gamma<4$ 
the exponents $\beta$ and $\delta$ read now $\delta=1+\beta$ and
$\delta=\beta(\gamma-2)$, so that, solving for $\beta$:
\begin{equation}
\beta=\frac{1}{\gamma-3}, 
\label{beta}
\end{equation}
as found in Refs. \cite{hav2} and \cite{Hav}.
Now the corresponding equation reads
\begin{equation}
 0=\langle\mu^{-1}\rangle\lambda x\partial_x^2\tilde{\phi}(x)
-x+\tilde{\phi}(x)+
c\Gamma(2-\gamma)\tilde{\phi}^{\gamma-2}(x).  
\label{scale2}
\end{equation}  
We note that $\beta$ is related to the 
probability $P_\infty\sim\Delta^\beta$ that a vertex belongs 
to the percolating cluster.
From Eqs. (\ref{scale1}) and (\ref{scale2}) for the
order parameter field $\tilde{\phi}$ we obtain a scaling relation for 
$R(\lambda)$
by inserting Eq. (\ref{scalean}) into Eq. (\ref{Rend}).  
To this end we subtract the pole $\rho_0/\lambda$ from
$R(\lambda)$ and expand in powers
of $\Delta$:
\begin{eqnarray}
\lefteqn{R(\lambda)-\frac{\rho_0}{\lambda}}
\nonumber\\
&=&-\frac{1}{\lambda}
\int_0^\infty dx\, e^{-x}\left\{G_0(\phi(x))-G_0(\phi_0(x))\right\}
\nonumber\\&\simeq&-\frac{\Delta^{1+\beta}}{\lambda}
\int_0^\infty dx\, e^{-\Delta^{1+\beta}x}
\nonumber\\&&
\times\big\{G_0(1-\Delta^{\beta}\tilde{\phi}(x,\lambda/\Delta^{2+\beta}))
-G_0(1-\Delta^{\beta}\tilde{\phi}_0(x))\big\}
\nonumber\\
&\simeq&
\Delta^{\beta-1}\frac{\langle k\rangle}{\lambda/\Delta^{2+\beta}}
\int_0^\infty dx\,\left\{\tilde{\phi}(x,\lambda/\Delta^{2+\beta})-\tilde{\phi}_0(x)\right\},
\label{rscale}
\end{eqnarray}
where $\tilde{\phi}_0(x)=
\lim_{\Delta\to0}\Delta^{-\beta}\{\phi_0(x\Delta^{1+\beta})-1\}$. 
Thus we have shown that $\rho_+(\lambda)$ 
obeys for $\lambda\sim\Delta$ close to $0$ a scaling law of the form  
\begin{equation}
  \rho_+(\lambda,\Delta)\simeq \Delta^{\beta-1}
\tilde{\rho}(\lambda/\Delta^{2+\beta}).\hspace{1cm}
\label{rhoscale}
\end{equation}
Furthermore, the scaling function $\tilde{\rho}(x)$ can be  
determined via Eqs. (\ref{rscale}), (\ref{scale1}), (\ref{scale2}), 
and (\ref{densres}).
From the preceding considerations it follows immediately that the shape of 
$\tilde{\rho}(\lambda)$ differs in the region $\gamma>4$ from
its shape in the region $3<\gamma<4$.
In the first region Eq. (\ref{scale1}) is valid 
and $\tilde{\rho}(\lambda)$ does not depend on $\gamma$, 
whereas in the second region the $\gamma$-dependent 
Eq. (\ref{scale2}) holds.

\section{Integration for scale free degree distributions and special 
distributions of bond strengths}
As shown in Ref. \cite{BAHZ01}, the analytical work simplifies considerably 
for the following distribution of bond strengths:
\begin{equation}
D(\mu )=\frac{1}{\mu^2}\exp (-1/\mu),  \label{31}
\end{equation}
since then the operator $\hat{{\bf O}}$, Eq. (\ref{21}), takes the form
\begin{equation}
  \label{simpop}
\hat{{\bf O}}=
\int_0^\infty d\mu \frac{1}{\mu^2}\exp
(-1/\mu)\exp\left[-\frac{\lambda}{\mu}x\partial_x^2\right]=\left[1+\lambda
  x\partial_x^2\right]^{-1}  
\end{equation}
For instance, applying $1+\lambda x\partial_x^2=\hat{{\bf O}}^{-1}$ to
both sides of
Eq. (\ref{39}), one obtains the ordinary second order differential equation 
\begin{equation}
\phi (x)+\lambda x\partial _{x}^{2}\phi (x)=e^{-x}G_1[\phi(x)].  
\label{32}
\end{equation}
As noted in Ref. \cite{BAHZ01}, the particular choice of
$D(\mu)$, Eq. (\ref{31}), does not change much the small $\lambda$
behavior of $\rho(\lambda)$, given that in Eq. 
(\ref{31}) the probability for small coupling
strengths $\mu$ is exponentially small.
In particular, $D(\mu)$ does not change the form of the 
function $\tilde{\rho}(\lambda)$, as  
only the first inverse moment $\langle \mu^{-1} \rangle$ 
enters Eqs. (\ref{scale1}) and (\ref{scale2}).
Eq. (\ref{32}) has to be solved subject to the boundary conditions
\begin{equation}
\phi(0)=1\;\;\mbox{and}\;\;\phi(\infty )=0.  
\label{35}
\end{equation}
In the limit $\lambda\to 0$ 
Eq. (\ref{32}) can be linearized around the first term $\phi_0(x)$ 
of the asymptotic expansion, Eq. (\ref{ansatz}).
This is achieved by inserting $\phi(x)=\phi_0(x)+\phi_l(x)$ into 
Eq. (\ref{32}) and keeping only linear terms in $\phi_l$, 
since from Eq. (\ref{ansatz}) we have $\phi_l(x)=O(\lambda)$. 
This results in the inhomogeneous linear equation
\begin{equation}
\lambda x\partial _{x}^{2}\phi_l(x)+
\left\{1-e^{-x}G_1'[\phi_0(x)]\right\}\phi_l(x)=
-\lambda x\partial _{x}^{2}\phi_0(x).  
\label{linear}
\end{equation}

To investigate specific scale-free degree distributions we
choose the following generating function 
\begin{eqnarray}
 G_0(\phi)&=&
\phi+\frac{1}{2}(1-\Delta)(1-\phi)^2-\frac{1}{9}\frac{\gamma-3}{\gamma-4}
(1-\phi)^3
\nonumber\\&&
+\frac{2}{3}\frac{1}{(\gamma-4)(\gamma-2)(\gamma-1)}(1-\phi)^{\gamma-1}.
\label{G0}
\end{eqnarray}
This form corresponds indeed to  a degree distribution $p_k$ which
obeys Eq. (\ref{scalefree}); the $c$-value in Eq. (\ref{scalefree}) is
\begin{equation}
c=\frac{2}{3}\frac{1}{(\gamma-4)(\gamma-2)(\gamma-1)\Gamma(1-\gamma)}.     
\label{c}
\end{equation}
The algebraic choice of $G_0(\phi)$ given by Eq. (\ref{G0})  
reduces considerably the effort needed to integrate Eq. (\ref{32}). 
Furthermore, Eq. (\ref{G0}) can be used in the whole interval
$3<\gamma<5$ containing the value $\gamma=4$ above which regular
mean-field exponents of percolation appear.
On the other hand, for values of $\gamma$ outside the interval $3<\gamma<5$ 
not all expansion coefficients of $G_0(\phi)$ are non-negative
and thus they cannot be viewed anymore as probabilities.  
Note that the poles in $4-\gamma$ of the  
last two terms in  Eq. (\ref{G0}) cancel and expanding Eq. (\ref{G0})
in powers of $4-\gamma$ we obtain for $\gamma=4$    
a branching point of the form $(1-\phi)^3\log(1-\phi)$ at $\phi=1$. 
Furthermore, from Eq. (\ref{rho0}) it follows that  $G_0(\phi)$ of 
Eq. (\ref{G0}) leads to $\rho_+(\lambda)$ being normalized to:
\begin{equation}
  \label{norm}
\int_0^\infty d\lambda\,\rho_+(\lambda)=\frac{1}{2}.  
\end{equation}

\subsection{Numerical procedure}
To numerically calculate the eigenvalue spectra of scale-free networks
we have performed extensive numerical diagonalizations of the 
Laplacians of these structures.

We create our structures by the recursive scheme introduced in the 
second section:
For each realization of the structure we first begin with an initial 
vertex, whose functionality $k$ is determined 
according to the probabilities $p_k$ derived from
the generating function Eq. (\ref{G0}). 
At the open end of each bond a new $k'$-functional vertex is 
placed, now with $k'$ distributed according to the probability
distribution $q_{k'}$
given in Eq. (\ref{verbindprob}); the latter procedure is then applied 
recursively to the $k'-1$ open bonds of this new vertex. The recursion
stops when no open bonds are left, i.e. when all outer vertices have 
functionality $k'=1$. Note, however, that due to the limited time and memory 
resources available for the subsequent diagonalizations, the total number of 
bonds has to be restricted to some maximum value $N_{\max }$. 
If a given recursion has not stopped before reaching a total of $N_{\max }$ 
bonds we proceed by closing all remaining open ends by a  $k'=1$-vertex 
and evaluate the properties of this truncated structure. 
Obviously, this also limits the range of validity of the resulting spectrum.
As observed in our previous study \cite{JFB03}, in the region
affected by the truncation the spectrum 
shows characteristic oscillations.
To verify this procedure
for the truncation limit  $N_{\max }=500$ used in general in this study,
we have also performed for some of the curves shown in Fig.4 additional
diagonalizations using  $N_{\max }=4000$.
The so obtained numerical results for $\log_{10}\rho(\lambda)$
agree within the symbol size for the whole range covered by the
$N_{\max}=500$ data shown in Fig.4 and in fact extend the
range of agreement with the solution of the differential Eq. 
(\ref{32}) by one order of magnitude.

For the distribution of bond strength 
of a given structure we chose either fixed bond strengths
$\mu=1$  or strengths $\mu$ distributed according to
Eq. (\ref{31}).  The connectivity
matrix ${\bf A}_{ij}$ with entries weighted by these 
$\mu $ is, by construction, a real, symmetric matrix. For all these
matrices we obtained the
eigenvalues using a combination of the Householder method and of the
tridiagonal QL diagonalization algorithm \cite{NR,NRa}.

We accumulated the eigenvalues of all structures generated for 
specific values of the parameters $\gamma$ and $\Delta$, where 
eigenvalues stemming from
a structure with $\left| S\right| $ monomers are weighted with a factor of $%
1/\left| S\right| $, as given
by Eqs. (\ref{averagedef}) and (\ref{rhos}).
For each of the data sets shown later in Figs 1, 3, and 4 the total
number of structures truncated
at $N_{\max}=500$ was $5\cdot 10^7$ and  for structures 
truncated at $N_{\max}=4000$ was $4\cdot 10^5$.

\section{Results}
\begin{figure}
\includegraphics[width=240pt]{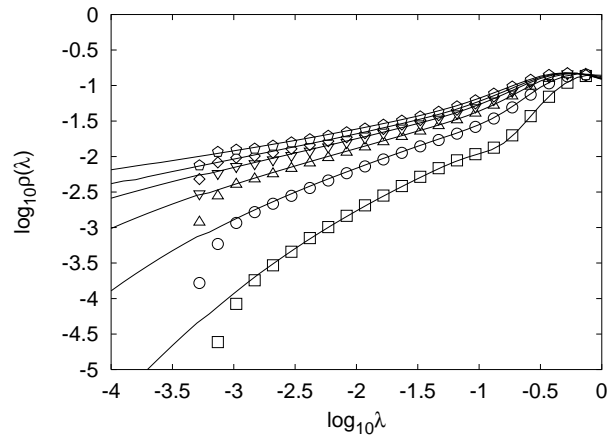}
\caption{\label{fig1} Density of eigenvalues 
  $\rho(\lambda)$ in double logarithmic scales. Displayed are spectra
  for the scale free degree distribution generated by $G_0(\phi)$, Eq.
  (\protect\ref{G0}), and random coupling strengths $\mu$ obeying Eq.
  (\protect\ref{31}). Here, $\gamma=3.5$ is fixed and $\Delta$ is 
  $0 (\rm \pentagon)$, 
  $0.02 (\lozenge)$,
  $0.04 (\triangledown)$,
  $0.08 (\vartriangle)$,
  $0.16 ({\scriptstyle\bigcirc})$,
  and $0.32 (\square)$ from above.
  Lines: numerical solution of Eq. (\protect\ref{32}). Symbols: direct
  diagonalization of randomly created structures.}
\end{figure}
In Fig. \ref{fig1} we display first the density of eigenvalues 
$\rho(\lambda)$ for the degree distribution $p_k$ generated 
by $G_0(\phi)$, Eq. (\ref{G0}), with $\gamma=3.5$ and for various values of 
$\Delta\le 0$. 
The random coupling strengths $\mu$ obey the distribution of Eq. (\ref{31}).
We obtained $\rho(\lambda)$ both through the numerical integration of 
Eq. (\ref{32}) and also through the direct 
numerical diagonalization of many structure realizations, as described above.
As can be seen, the $\rho(\lambda)$ obtained by the two methods agree very 
well with each other over a large $\lambda$-range,
thus supporting our theoretical considerations.
The deviations of the curves from each other 
for small $\lambda$ are due to the 
limitations imposed by our direct diagonalization approach; in fact
the sharp decay of the numerical results for $\lambda<10^{-3}$ is an artefact.
The curves of Fig. \ref{fig1} possess a shoulder at 
$\log_{10}\lambda\simeq -0.7$ which is most evident for the lowest curve 
corresponding to $\Delta=0.32$.
However, this structure is caused by the choice 
$G_0(\phi)$, Eq. (\ref{G0}), and is not specific for scale-free degree
distributions. As discussed above, scale-free networks are characterized 
by the behavior for {\em small values} of $\lambda$. 
\begin{figure}
\includegraphics[width=240pt]{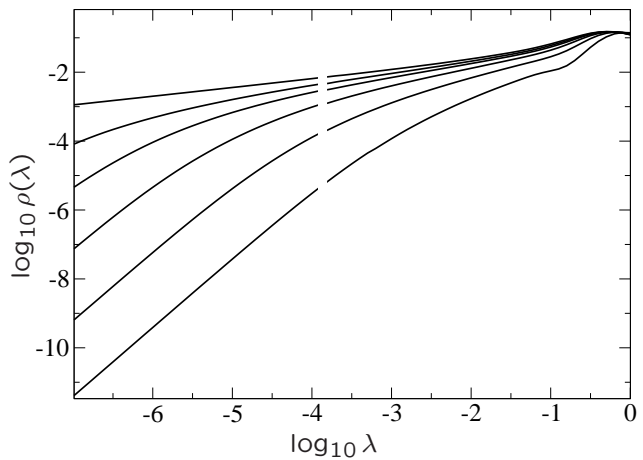}
\caption{\label{fig1b} Density of eigenvalues 
  $\rho(\lambda)$, obtained from the integration
  of Eq. (\ref{32}) (right hand side of the figure) and from Eq. 
(\ref{linear}) 
(left hand side) for the same parameter values as in Fig \ref{fig1}.
To render this difference clear, we have left out a small region
  around $\log_{10}\lambda=-3.8$.}
\end{figure}
To investigate the small $\lambda$ behavior, we show
in Fig. \ref{fig1b} $\rho(\lambda)$ for 
the same values of the parameters $\gamma$ and $\Delta$, but extending to 
much smaller values of $\lambda$.   
In this plot the results below $\lambda<10^{-3.8}$ 
are obtained from the linearized Eq. (\ref{linear}), an 
approximation which we expect to be exact in the
limit $\lambda\to0$. In fact, at $\lambda=10^{-3.8}$, where the
approximation and the exact curve get together,
the relative error of $\rho(\lambda)$ 
amounts to about $1\%$, which is already hard to 
observe in the plots.   
For  $\Delta$ not too close to the percolation threshold 
at $\Delta=0$ and for small $\lambda$ 
the slopes of the curves in the double logarithmic
plot of Fig. \ref{fig1b} 
tend to a constant. This would imply a simple algebraic dependence: 
\begin{equation}
\rho(\lambda)\sim c(\Delta)\lambda^{\alpha_1},\;\;\mbox{for}\;\;
\lambda \to 0, \label{decay}
\end{equation}
with a positive exponent $\alpha_1$ and a $\Delta$-dependent
coefficient $c(\Delta)$.
This differs from the situation expected to hold on 
classical random graphs with sufficiently fast
decaying degree distributions $p_k$, where heuristic
arguments have been given \cite{Br88,BAHZ01} for the existence
of Lifshitz tails in the density of states. 
In the latter situation one should observe the form \cite{JFB03}
\begin{equation}
  \label{liftail}
\rho(\lambda)\sim 
\exp\left[-\frac{A(\Delta)}{\sqrt{\lambda}}\right],\;\;\mbox{for}\;\; 
\lambda\to 0,  
\end{equation}
where $ A(\Delta) \sim \Delta^{3/2}$ for $\Delta\to0$.  
This behavior stems from the fact that small eigenvalues are produced
by large, quasi linear regions, which however, occur with very 
small probability.
Since for scale free degree
distributions such linear regions are not likely to
occur, there have to be other types of configurations which lead to an
increase in the occurrence of small eigenvalues. 
For instance two vertices, each of very large degree, moving against each
other produce a very low eigenvalue.

At the percolation threshold $\Delta=0$ we infer from Fig. 2 for small
$\lambda$ an algebraic decay 
of the form of Eq. (\ref{decay}), with an exponent $\alpha_2$, which
however differs from $\alpha_1$. One has namely $\alpha_2<\alpha_1$. 
Thus close to $\Delta=0$
we encounter here a crossover behavior between two algebraic decays with
different powers $\alpha_1$ and $\alpha_2$.
The scaling Eq. (\ref{rhoscale}) suggests that this crossover
should take place at $\lambda\sim\Delta^{2+\beta}$.

\begin{figure}
\includegraphics[width=240pt]{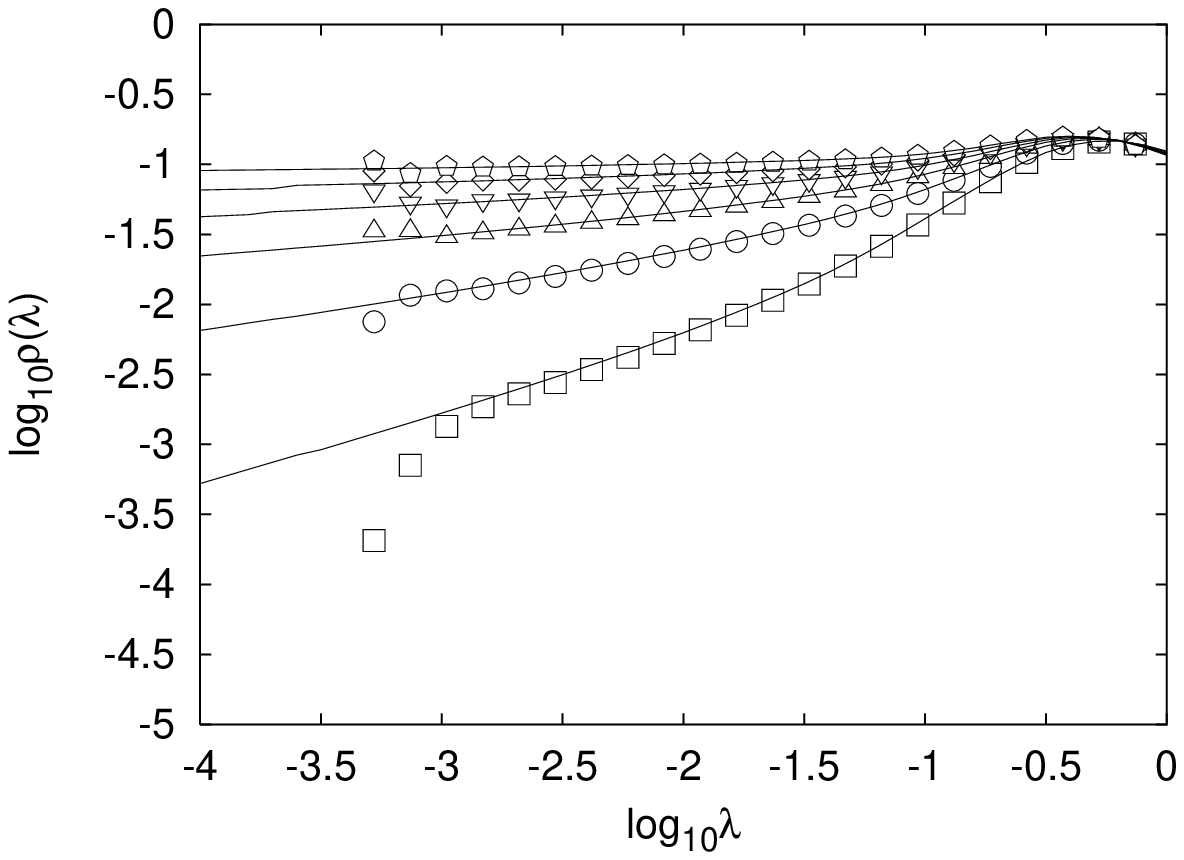}
\caption{\label{fig2} Density of eigenvalues $\rho(\lambda)$  in double
  logarithmic scales at the percolation threshold $\Delta=0$ for
  $G_0(\phi)$, Eq. (\ref{G0}) and $\mu$ obeying Eq. (\ref{31}). Here
  $\gamma$ is varied, being taken to be $\gamma=4.5 (\pentagon)$,
  $4.25 (\lozenge)$, $4 (\triangledown)$, $3.75 (\vartriangle)$, $3.5
  ({\scriptstyle\bigcirc})$ and $3.25 (\square)$ from above. Lines:
  integration of Eq. (\ref{32}), Symbols: direct diagonalization.}
\end{figure}
\begin{figure}
\includegraphics[width=240pt]{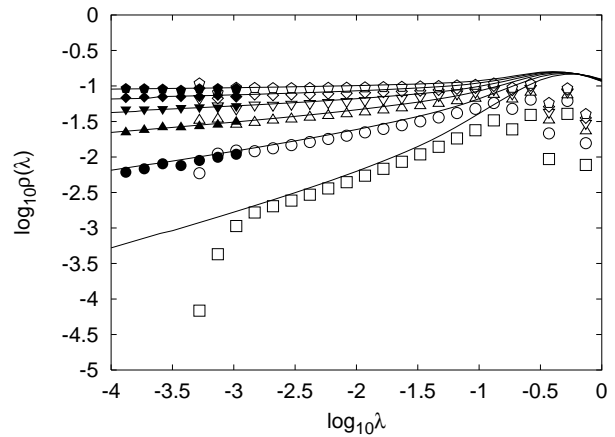}
\caption{\label{fig2b} Density of eigenvalues $\rho(\lambda)$ 
at the percolation threshold $\Delta=0$ for fixed
  coupling strengths, $\mu=1$ in comparison with the 
analytical results for $\mu$ obeying Eq. (\ref{31}) (straight lines). 
The symbols show simulation data truncated at
$N_{\max }=500$ (open symbols) and $N_{\max }=4000$ (filled symbols).
The values of $\gamma$ and $\Delta$ and the 
symbol shapes are as in Fig.\ref{fig2}}
\end{figure}
To determine the $\gamma$ dependence of $\alpha_2$,  
we display in Fig. \ref{fig2} $\rho(\lambda)$ in double
logarithmic scales for $\Delta=0$ and for various values of $\gamma$. 
Assuming that for 
$\lambda\to 0$ the slopes of the plotted curves tend to a constant, say
$\alpha_2$  we infer  for 
$\gamma=4.5,\;4.25,\;4,\;3.75,\;3.5$ and $3.25$ the values
$\alpha_2=0.017,\;0.015,\;0.04,\;0.113,\;0.25$ and $0.5$, respectively. 
For $\gamma>4$ we expect to encounter the classical mean-field scaling
function;  since then $\rho(\lambda)$ tends
to a constant for $\lambda\to0$ at $\Delta=0$ \cite{JFB03,BAHZ01}, 
this implies that $\alpha_2=0$ which is in good
agreement with the first two values $\alpha_2=0.017$ for $\gamma=4.5$ 
and $\alpha_2=0.015$ for $\gamma=4.25$. Moreover,  
directly at $\gamma=4$, we expect possible logarithmic 
corrections to the power law, Eq. (\ref{decay}) which explains the
value $\alpha_2=0.04$ which is slightly too large.
For $3<\gamma<4$ it 
turns out that we can reproduce the $\gamma$-dependence of $\alpha_2$
through the relation:
\begin{equation}
\alpha_2=\frac{4-\gamma}{2\gamma-5}.  
\label{guess}
\end{equation}
In a similar way, we obtain 
from the small $\lambda$ behavior of 
$\rho(\lambda)$ for $\Delta>0$ that $\alpha_1=2\gamma-5$ holds.

The $\gamma$ dependence of $\alpha_2$ can be derived
from the scaling relation Eq. (\ref{rhoscale}) by using the fact that
the behavior of $\tilde{\rho}(\lambda)$ for $\lambda\gg 1$ is given by
the algebraic dependence of $\rho(\lambda)$ for $\lambda
\ll1$ at $\Delta=0$
\begin{eqnarray}
\tilde{\rho}(\lambda)&=&\lim_{\Delta\to 0}
\Delta^{1-\beta}\rho(\lambda\Delta^{2+\beta},\Delta)\sim 
\Delta^{1-\beta}(\lambda\Delta^{2+\beta})^{\alpha_2}
\nonumber\\&=&
\Delta^{1-\beta+\alpha_2(2+\beta)}\lambda^{\alpha_2}.
\end{eqnarray}
For this to give a reasonable limit the exponent
$1-\beta+\alpha_2(2+\beta)=0$ of $\Delta$ has to vanish and solving
this equation for $\alpha_2$ proves the relation
Eq. (\ref{guess}).  
Similarly, the behavior of $\tilde{\rho}(\lambda)$ for $\lambda\ll 1$ can be
derived from the algebraic dependence Eq. (\ref{decay}) of  
$\rho(\lambda)$ for $\lambda\ll1$ and $\Delta>0$
\begin{eqnarray}
\tilde{\rho}(\lambda)&=&\lim_{\Delta\to 0}
\Delta^{1-\beta}\rho(\lambda\Delta^{2+\beta},\Delta)\sim 
\Delta^{1-\beta}c(\Delta)(\lambda\Delta^{2+\beta})^{\alpha_1}
\nonumber\\&=&
\Delta^{1-\beta+\alpha_1(2+\beta)}c(\Delta)\lambda^{\alpha_1}.
\end{eqnarray}
By the same arguments as above this leads to 
$c(\Delta)\sim \Delta^{\beta-1-\alpha_1(2+\beta)}$, but does not fix
the exponent $\alpha_1$.

As shown by Eqs. (\ref{scale2}) and (\ref{rscale}), the scaling function
$\tilde{\rho}(\lambda)$ of Eq. (\ref{rhoscale}) should be 
model independent and therefore 
should not depend on the particular choice of $D(\mu)$, Eq. (\ref{31}). 
In particular, at $\Delta=0$ a power law with the  
exponent $\alpha_2$ should still hold.
This is corroborated by 
Fig. \ref{fig2b}, in which we display the density of eigenvalues  
$\rho(\lambda)$ for fixed bond
strengths, $\mu=1$, obtained from the direct diagonalization of random
matrices,
together with the analytical results for the distribution, Eq. (\ref{31}), of 
coupling strengths.

\begin{figure}
\includegraphics[width=240pt]{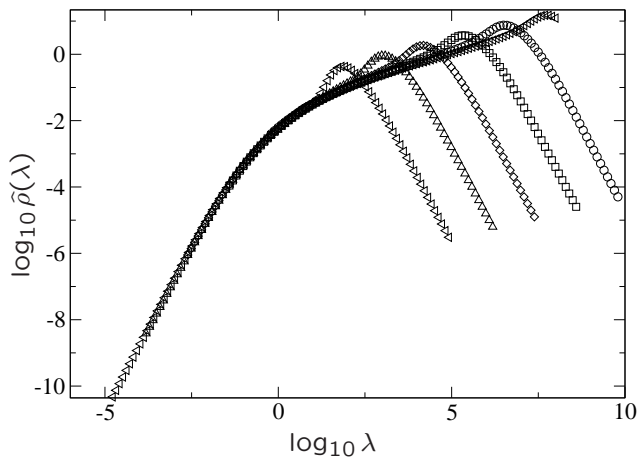}
\caption{\label{fig3} The quantity
  $\hat{\rho}(\lambda,\Delta)=\Delta^{-1}\rho(\lambda\Delta^4)$,
  for $\gamma=3.5$ in double logarithmic scales. The different curves
  correspond to $\Delta=0.01,0.02,0.04,0.08$ and $0.16$ from the right.}
\end{figure}
To investigate the range of validity of the scaling law
Eq. (\ref{rhoscale}) we display in double logarithmic
scales for fixed $\gamma=3.5$ in Fig. 5 and $\gamma=4.5$ in Fig. 6 the
quantity
$\hat{\rho}(\lambda,\Delta)\equiv
\Delta^{1-\beta}\rho(\lambda\Delta^{2+\beta},\Delta)$ 
which tends for $\Delta\to0$ to the scaling function
$\tilde{\rho}(\lambda)$ of Eq. (\ref{rhoscale}).
In both figures we show curves for various values of $\Delta$ close
to $\Delta=0$. 
For  $\lambda$ and $\Delta$ small enough  
the curves for $\hat{\rho}(\lambda,\Delta)$ should collapse into a single
one, given by the scaling function
$\tilde{\rho}(\lambda)$ of Eq. (\ref{rhoscale}). In Fig. 5 the
collapse appears 
roughly for $\lambda\Delta^4<10^{-2}$ and in Fig. 6 for 
$\lambda\Delta^3<10^{-2}$.  
In Figs. 5 and 6 the scaling functions
$\tilde{\rho}(\lambda)$ are given by the envelopes of the curves   
$\hat{\rho}(\lambda,\Delta)$ for different values of $\Delta$.
These envelopes $\tilde{\rho}(\lambda)$ seem to be monotonic
growing functions.
In Fig. 5 $\tilde{\rho}(\lambda)$
shows for the limiting case $\lambda\to 0$  
the algebraic behavior $\tilde{\rho}(\lambda)\sim\lambda^{\alpha_1}$
but in Fig. 6 it is not possible to observe any 
algebraic dependence.

\begin{figure}
\includegraphics[width=240pt]{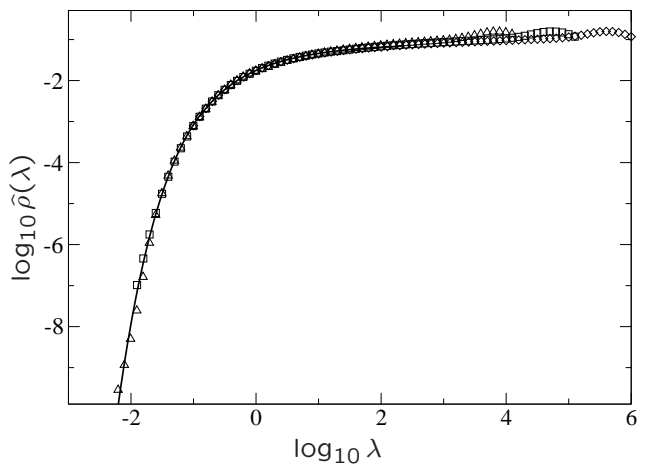}
\caption{\label{fig3b}
The quantity $\hat{\rho}(\lambda,\Delta)=\rho(\lambda\Delta^{3})$ 
for $\gamma=4.5$
in double logarithmic scales. We show curves for 
$\Delta=0.01$ (diamonds), $\Delta=0.02$ (squares) and $\Delta=0.04$
(triangles). In addition, we display as a solid line 
the exact scaling function obtained
by solving Eq. (\ref{scale1}).}
\end{figure}
In Fig. 6, the function 
$\tilde{\rho}(\lambda)$ given as a solid line 
is obtained from the direct integration
of Eq. (\ref{scale1}), which scales. More generally,
for $\gamma>4$ we obtain $\beta=1$; we note that for $\gamma>4$ all
network ensembles lead to the same scaling function 
$\tilde{\rho}(\lambda)$.
Note that the derivation of Eqs. (\ref{scale1}) and (\ref{rscale}) in 
Sec. IV shows that the essential 
condition for this scaling function to hold is that the 
degree distribution $p_k$ decay faster than $k^{-4}$.
This condition is certainly fulfilled for classical random graphs, like
the bond diluted Cayley tree or the Erd\"os-R\'enyi random graph.

\section{Conclusions}
In this work we investigated the eigenvalues of Laplacians 
of structures belonging to a general type of
tree-like networks, in which the  vertex
degrees are randomly distributed.
The Laplacian is of special interest, since it
determines several, very important dynamic quantities associated with the
network.
For degree distributions $p_k$ with a power law
tail, $p_k\sim k^{-\gamma}$, we obtained $\rho(\lambda)$, 
the ensemble averaged density
of eigenvalues, based on two different methods. 
First, in a traditional way, by performing numerical 
diagonalization techniques \cite{GKK,FDBV}; 
second, using the replica method of statistical physics.
The second approach allows to evaluate the ensemble averaged
$\rho(\lambda)$ based on an 
analytical integral equation.
For large $\lambda$-domains it turns out that the
agreement between the results obtained by the two methods is very good.

Of special interest is the behavior of $\rho(\lambda)$ 
close to the percolation threshold. Here an
infinite cluster appears, and it is known that 
the exponent $\gamma$ which governs the large $k$ behavior of $p_k$ 
affects the critical exponents of the percolation problem 
\cite{Hav}.  
With the help of our integral equation approach we were able to study the 
scaling properties of $\rho(\lambda)$ close to the percolation
threshold and to determine numerically the corresponding, $\gamma$-dependent 
scaling functions. In agreement with Ref. \cite{Hav}, we find that in
the region $\gamma>4$ one recovers
the critical properties of classical random graphs.

The long time dynamics is governed by the small $\lambda$ behavior of 
$\rho(\lambda)$. For this we found two algebraic forms 
$\rho(\lambda)\sim \lambda^{\alpha_1}$ and 
$\rho(\lambda)\sim \lambda^{\alpha_2}$, where the first relation holds
{\em below}
and the second {\em at} 
the percolation threshold. On the basis of the numerical
results of the integral equation we conjecture that
$\alpha_1=2\gamma-5$ and $\alpha_2=(4-\gamma)/\alpha_1$ hold. 
 We find that in scale-free networks 
very small eigenvalues occur with higher probability than in classical 
random graphs. We conjecture that this
finding is due 
to the existence of highly connected vertices.  
 
\acknowledgments
The support of the DFG, of the Fonds der Chemischen Industrie and of
the BMBF are gratefully acknowledged.
We are much indebted to Profs. S. Havlin and Y. Holovatch for
enlightening discussions.

%\bibliography{jasch}

\end{document}